\documentclass[aps,pra,twocolumn,showpacs,superscriptaddress,footinbib,groupedaddress,longbibliography]{revtex4-1}
\usepackage{graphicx}  
\usepackage{dcolumn}   
\usepackage{bm}        
\usepackage{amssymb}   
\usepackage{gensymb}   
\usepackage{amsmath}
\usepackage{float}
\usepackage{xcolor}
\usepackage[english]{babel}
\usepackage{color}
\usepackage{ulem}
\usepackage{soul}	

\usepackage[colorlinks=true,citecolor=black,urlcolor=blue,linkcolor=black]{hyperref} 

\newcommand{\keff}{k_{\text{eff}}}
\newcommand{\er}{\vec{e}_r}
\newcommand{\etheta}{\vec{e}_{\theta}}
\newcommand{\poskeff}{\frac{\hbar\keff T}{m}}
\newcommand{\phikeff}{\frac{\hbar\keff^2 T}{m}}
\newcommand\stxt[1]{_{\text{#1}}} 

\begin{document}

\title{Accurate trajectory alignment  in cold-atom interferometers with  separated laser beams}

\author{M. Altorio, L. A. Sidorenkov, R. Gautier, D. Savoie, A. Landragin and R. Geiger}%
\date{\today}%
\email{remi.geiger@obspm.fr}
\affiliation{LNE-SYRTE, Observatoire de Paris-Universit\'e PSL, CNRS, Sorbonne Universit\'e,  61 avenue de l'Observatoire, 75014 Paris, France.}

\begin{abstract}
Cold-atom interferometers commonly face systematic effects originating from the coupling between the trajectory of the atomic wave packet and the wave front of the laser beams driving the interferometer. 
Detrimental for the accuracy and the stability of such inertial sensors, these systematics are particularly enhanced in architectures based on spatially separated laser beams. Here we analyze the effect of a coupling between the relative alignment of two separated laser beams and the trajectory of the atomic wave packet in a four-light-pulse cold-atom gyroscope operated in fountain configuration. We present a method to align the two laser beams at the 0.2~$\mu$rad level \emph{and} to determine the optimal mean velocity of the atomic wave packet with an accuracy of 0.2~$\textrm{mm}\cdot\textrm{s}^{-1}$. 
Such fine tuning constrains the associated gyroscope bias to a level of $1\times 10^{-10}~\textrm{rad}\cdot\textrm{s}^{-1}$.
In addition, we reveal this coupling using the point-source interferometry technique by  analyzing single-shot time-of-flight fluorescence traces, which allows us to measure large angular misalignments between the interrogation beams.
The alignment method which we present here can be employed in other sensor configurations and is particularly relevant to emerging gravitational wave detector concepts based on cold-atom interferometry.

\end{abstract}

\maketitle

\section{Introduction}
\label{sec:intro}
Cold-atom inertial sensors based on light-pulse interferometry  are  developed by several groups in the world for various applications, such as gravimetry and gradiometry, metrology, tests of fundamental physics, navigation and gravitational wave astronomy.
In these sensors, atomic phase shifts generated by inertial forces originate from the relative motion between the free falling atoms and the local reference frame represented by the optical phase of the lasers driving the beam splitters and mirrors for the atomic waves.
As the laser beams are not  perfect plane waves,  the sampling of the inhomogeneous wavefronts by the finite size of the atom cloud at each pulse results in a systematic shift. Changes in the mean atomic trajectory or temperature then lead to a limitation of the stability of the sensors.
Wavefront aberrations coupled to the transverse expansion of the atom cloud in the laser beam is, for example, a limiting factor to the accuracy of cold-atom gravimeters \cite{LouchetChauvet2011,Schkolnik2015,Wang2018} and has pointed towards using ultra-cold atom sources for improved accuracy \cite{Karcher2018}.
Even in differential configurations such as used in atomic gyroscopes with counter-propagating atom clouds \cite{Fils2005,Gauguet2009,Tackmann2012,Yao2018}, in gravity gradiometry \cite{Trimeche2019} or gravitational wave detectors \cite{Hogan2011}, stochastic variations of the atom trajectories or of the laser field wavefront pose severe  constraints on the optics and on the temperature and initial position jitter of the atom source.

Such effects are even more pronounced when the atom interferometer is operated with laser beams that propagate perpendicularly to the atom velocity in order to open a physical area in the interferometer \cite{Dutta2016}. This problem was first identified in Ref.~\cite{Fils2005} and investigated in the case of a dual cold-atom source gyroscope in Ref.~\cite{Gauguet2009}.
When using separated laser beams, a systematic shift occurs even with a plane wavefront as soon as different beams are not perfectly aligned. The systematic shift associated with such angular misalignement scales with the interrogation time $T$ (time between light-pulses) and the  atom initial velocity, while the inertial signal scales with the area enclosed by the interferometer paths (scaling with $T^2$ ($T^3$) in a three (four) pulse gyroscope).
A method to align with $\mu$rad precision the 3 laser beams in a Mach-Zehnder-like configuration of atom  interferometer gyroscope  was  presented in Ref.~\cite{Tackmann2012} for a total interrogation time $2T\simeq 50$~ms   (this method was later used in Ref.~\cite{Yao2018} where $2T=104$~ms). However, the  residual systematic shift was not evaluated in this study.
We report here on a method to align the interrogation beams \emph{and} to find the optimal atom trajectory, which allows us to give an upper bound on the residual systematic shift that becomes of second order in the small parameters of the problem. While this alignment method is general to several sensor architectures, we illustrate it in the case of a 4-light pulse interferometer geometry where the atom cloud is launched in a fountain configuration with a total interrogation time of 800~ms, which has shown favorable performance compared to 3-light pulse cold-atom gyroscopes \cite{Dutta2016,Savoie2018}.

The article is organized as follows: section~\ref{sec:effect} presents the effect of the coupling of the atom initial velocity to the relative alignment between the two interrogation beams; section \ref{sec:expt} presents the experiment and the main methods used in this work; section \ref{sec:results} shows the measurements of the contrast loss and of the phase shifts derived in section \ref{sec:effect}; section \ref{sec:point_source} presents an analysis of the effect using the technique of point source interferometry \cite{Dickerson2013,Hoth2016} providing a direct measurement of a velocity-dependent phase shift in a single time-of-flight fluorescence trace; we finally conclude in section \ref{sec:conclusion} and discuss the importance of the effect in other sensor configurations.

\section{Coupling of the  atom velocity to the  relative alignment between two interrogation beams}
\label{sec:effect}
\begin{figure}[b]
    \centering
    \includegraphics[width=\linewidth]{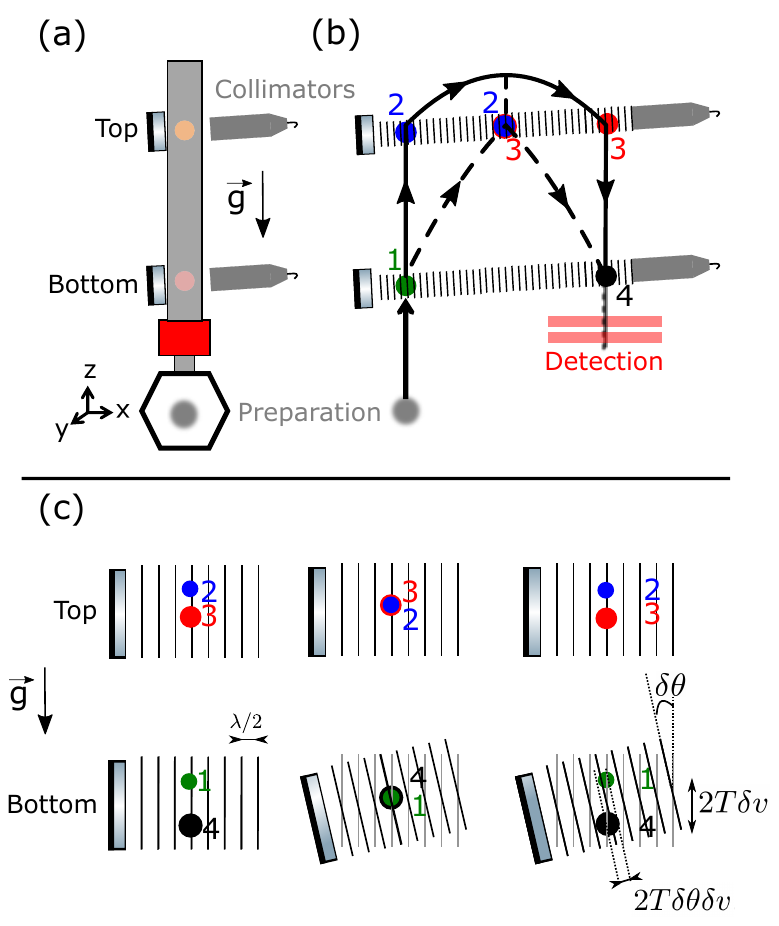}
    \caption{
    (a) Sketch of the fountain with the laser beam configuration.
    (b) Path of the atomic waves in the interferometer (in the $(xz)$ plane). The numbers $1,2,3,4$ refer to the indices of the light pulses at times $t=0,\frac{T}{2},\frac{3T}{2},2T$.
    (c)    Illustration of the systematic phase shift induced by the coupling of atom velocity to mirror misalignment,  in the vertical direction.
     No phase shift is present if: (left) the mirrors are parallel even if the initial velocity is not the optimal one (see text), or (middle)  the initial velocity is the optimal one even if the two mirrors are misaligned . In the right column, the mirrors are misaligned and the initial velocity is not the optimal one, resulting in a systematic phase shift $\Delta \Phi=2T\keff\delta \theta\delta v$. }
    \label{fig:sketch_full}
\end{figure}

We analyze the effect of the coupling between the initial velocity of an atom entering a 4-light pulse interferometer  and the relative alignment of the two laser beams realizing the beam splitters (bottom beam) and mirrors (top beam) for the atom wave. Full details about the 4-pulse geometry in fountain configuration are given in our previous works \cite{Dutta2016,Savoie2018}. A sketch of the laser configuration is recalled here in Fig.~\ref{fig:sketch_full}(a) showing the two beams separated by a distance  $3gT^2/8\simeq 0.59$~m ($T=0.4$~s). Panel (b) illustrates the path of the atomic waves in the interferometer.
Each beam carries two laser frequencies to drive stimulated Raman transitions and is retro-reflected by a mirror. The relative direction of the effective Raman wave vector between the bottom and top beams is given to first order by the  relative angle $\delta\theta$ between the two retro-mirrors.


At each light pulse the relative phase between the two Raman lasers is imprinted on the diffracted part of the atomic wave function.
The total phase shift between the two arms of the interferometer is \cite{Stockton2011,Dutta2016}
\begin{equation}\label{eq:general_deltaPhi}
\Delta\Phi = \vec{k}\stxt{B}\cdot\vec{r}\left(0\right) - 2\vec{k}\stxt{T}\cdot\vec{r}\left(\frac{T}{2}\right) + 2\vec{k}\stxt{T}\cdot\vec{r}\left(\frac{3T}{2}\right) - \vec{k}\stxt{B}\cdot\vec{r}\left(2T\right)
\end{equation}
where $\vec{r}(t)$ is the classical position of the center of mass of the wavepacket at time $t$, and $\vec{k}\stxt{B}$ ($\vec{k}\stxt{T}$) is the effective two-photon wave vector for the bottom (top) beam.
Denoting as $\delta\theta$ the angular mirrors' misalignment, we  express the two Raman wave-vectors as:
\begin{align}\label{eq:relation_k}
    \vec{k}\stxt{B} =& \keff\er\nonumber \\
    \vec{k}\stxt{T} =& \keff(\cos\delta\theta\er + \sin\delta\theta\etheta)
\end{align}
where $\keff = 4\pi/\lambda$ is the momentum transfer during Raman transition ($\lambda$ is the laser wavelength), $\er$ is the unitary vector in the direction of the bottom beam, $\etheta$ is the normal to $\er$.
Calculating the classical trajectory of the atom and inserting Eq.~\eqref{eq:relation_k} into Eq.~\eqref{eq:general_deltaPhi}, we obtain the expression for the phase shift induced by mirrors' misalignment, to the first order in $\delta\theta$ (see Appendix~\ref{sec:AppendixA} for the details of the derivation):
\begin{equation}\label{eq:velocity_theta_deltaPhi}
\Delta\Phi(\vec{v}, \delta\theta\etheta) = 2T\keff\delta\theta\etheta\cdot(\vec{v} + T\vec{g}),
\end{equation}
where $\vec{g}$ is the local gravity acceleration (sum of gravitational and centrifugal components) and $\vec{v}$ is the velocity of the atom at the time of the first pulse, referred as initial velocity hereafter.
We define the optimal velocity as $\vec{v}\stxt{opt} = - T\vec{g}$, which leads to a cancellation of the systematic shift. Hereafter we denote as $\delta \vec{v}=\vec{v}-\vec{v}\stxt{opt}$ the offset from the optimum.
The effect is two dimensional and we will decompose its two contributions in a vertical component (projection of $\etheta$ in the $(xz)$ plane) and an horizontal component (projection in the $(xy)$ plane).
In each direction (horizontal, vertical), the systematic shift amounts to 12~mrad of interferometer phase per $1 \ \mu$rad of mirror' misalignment and $\delta v=1$~mm.s$^{-1}$.
In the following, we  use this systematic shift as a tool to accurately align the atomic trajectory and minimize the angular misalignments of the mirrors.

\section{Experimental setup}
\label{sec:expt}

\subsection{Atomic fountain}
Details about the experimental apparatus are given in our previous works \cite{Dutta2016,Savoie2018}. We recall here the main elements for completeness.

We start our cycle by laser cooling and trapping about $10^7$ Cesium atoms in a  magneto-optical trap (MOT) loaded from a 2D-MOT. We then launch an atomic cloud vertically in a fountain configuration with a mean initial velocity of about 5.0~m.s$^{-1}$. The launching is followed by $ 2~\textrm{ms}$ of optical molasses, resulting in a velocity distribution $f(v)$ described by the so-called Lorentz-B function (see Appendix~\ref{sec:appendix_vel_distrib} for a detailed discussion on the choice of fitting function). In the following, we will approximately characterize the atomic cloud by effective temperature $T_{at}\simeq 1.8~\mu\textrm{K}$ corresponding to a Gaussian velocity distribution of standard deviation $\sigma_{v}\simeq 11~\textrm{mm.s}^{-1}$. 
After the molasses phase, the  atoms are selected in the state $|F=4, m_F=0\rangle$ with a pulse of magnetic field gradient and enter the interrogation region. The atoms interact with two laser frequencies driving stimulated Raman transitions coupling the two states $|F=3, m_F=0\rangle$ and $|F=4, m_F=0\rangle$.
The Raman beams enter the interrogation region from two collimators and are retro-reflected. The beams are tilted  by an angle $\theta_{0}=3.79(1)\degree$ with respect to horizontal direction (orthogonal to $\vec{g}$), in order to lift the degeneracy between the Raman transitions associated with $\pm\hbar\keff$ momentum transfer.


To fine-tune the relative alignment between the two retro-reflected beams, we use a  non-magnetic piezomotor-controlled mirror mount (SR200iNM-HS-200-2PZT assembled by Lioptec and Physik Instrumente (PI)) to hold the bottom mirror of 50.8~mm diameter. We control piezomotors driving either vertical or horizontal tilt of the mirror by steps with a typical size of 23~nm.  We calibrated the step size of our mirror by recording the position shift of the retro-reflected beam on a CCD-camera and obtained  angular variations of 0.39(2)~$\mu$rad/step and 0.38(2)~$\mu$rad/step for the vertical and horizontal directions, respectively \cite{footnoteMirror}.

\subsection{Detection system}
\label{subsec:detection_system}

At the end of the interferometric cycle, the atomic population in the output states $F=4$ and $F=3$ is detected by means of fluorescence detection.
The probability of the atom with initial velocity $\vec{v}$ to be in state $F=4$ at the output of the interferometer is given by
\begin{equation}
P_{F=4}\equiv P(\vec{v}, \delta\theta\etheta)=\frac{1}{2}+C\cos\big(\Delta\Phi(\vec{v}, \delta\theta\etheta) + \phi\big),
\label{eq:probability_to_phase_link}
\end{equation}
where $2C$ is the fringe visibility (assumed to be independent of atom velocity) and $\phi$ incorporates the constant phase shift induced by Earth rotation and phase noise contributions due to, for example, vibration noise or laser phase noise.
Accounting for the finite-temperature velocity distribution of the atoms, $f(\vec{v})$, the  probability to find an atom with initial velocity $\vec{v}$ in state $F=4$ is given by  $P(\vec{v}, \delta\theta\etheta)f(\vec{v})$.

\begin{figure}[t]
	\centering
	\includegraphics[width=\linewidth]{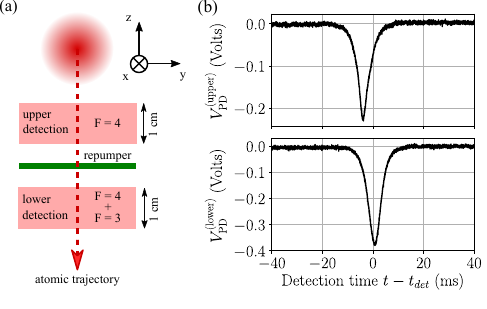}
	\caption{(a) Sketch of the detection scheme (to scale): atoms fall down into the detection region sequentially passing through the upper detection, middle repumping, and lower detection light sheets. The light scattered by the atoms within the detection regions is collected by upper and lower photodiodes resulting in signals shown in (b). The geometry of the detection optics ensures negligible contribution of the scattering from upper (lower) regions into the signal of lower(upper) photodiode.}
	\label{fig:Figure_detection_scheme}
\end{figure}

Fig.~\ref{fig:Figure_detection_scheme}  shows our detection scheme and typical atomic signals. We illuminate the atomic cloud with light resonant with the $F=4\rightarrow F^{'}=5$ transition for a duration of 80~ms and with an intensity $I\simeq 0.2I_{\rm{sat}}$, $I_{\rm{sat}}$ being the saturation intensity. The vertical size of each light sheet is $d = 10~\textrm{mm}$.
We record the fluorescence with a two-quadrant photodiode imaging the  upper and lower detection regions, as the atoms traverse them sequentially (the imaging magnification equals $0.3$). For the same time, a thin repumping lightsheet (2~mm in height) is on and is resonant with $F=3\rightarrow F^{'}=4$. The integrated signal in the upper light sheet is thus proportional to the number of atoms in $F=4$, while the integrated  signal in the lower light sheet is proportional to the total atom number. The total flight time from the launch to the center of the detection time window is $t_{det}=984~\textrm{ms}$. 

By design, our detection scheme does not discriminate the horizontal position of the atoms within the light sheets and therefore integrates over all velocity classes along the $x$- and $y$-directions. Moreover, integrating over time the  signal from the upper (lower) light sheet, $S_P(t)$,  ($S_N(t)$) integrates over the velocity distribution in the $z$-direction, yielding the total number of atoms in the $F=4$ internal state, $N_{F=4}$ (total number of atoms, $N$). In the next section, we determine the transition probability from these integrated signals as the ratio $N_{F=4}/N$.

\section{Results}
\label{sec:results}

\subsection{Alignment of the mirrors}

\begin{figure}
	\centering
	\includegraphics[width=\linewidth]{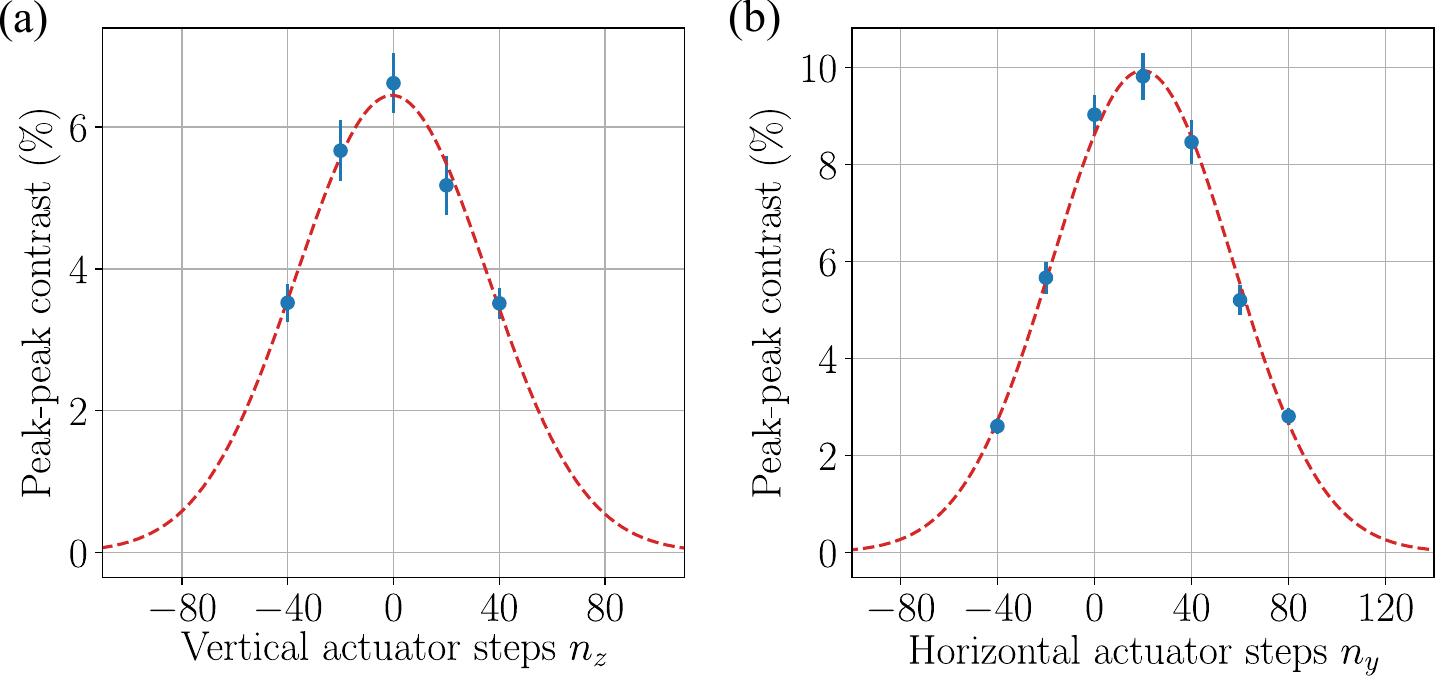}
	\caption{Contrast of the interferometer as a function of mirror steps in (a) vertical ($n_{z}$) and  (b) horizontal ($n_{y}$) directions. Error bars are standard deviations of fit-by-packet in a phase-scan acquisition (see text). Dashed red lines are phenomenological Gaussian fits to the data ($A_{z,y}\exp(-(n_{z,y}-n_{z0,y0})^{2}/2\sigma_{z,y}^{2})$ revealing  $n_{z0}=-0.6(4.0)$, $\sigma_{z}=36.2(6.1)$, $n_{y0}=19.8(2.3)$, $\sigma_{y}=37.1(2.4)$. The difference in maximum value of contrast for two directions is due to sequential optimization: first $z$-direction, then $y$-direction with $n_{z}=n_{z0}$. }
	\label{fig:contrast_ZY}
\end{figure}

A misalignment between the two mirrors translates into a velocity-dependent phase shift, which, after integration over velocities, results in a loss of interferometer visibility. The observation of this effect can be used to minimize the misalignment between the two mirrors \cite{Tackmann2012}.
We show in Fig.~\ref{fig:contrast_ZY} the visibility of the interferometer for various values of mirrors' relative misalignment in  the vertical ($\delta\theta_z$) and horizontal ($\delta\theta_y$) directions. To obtain the values of visibility, we varied the interferometer phase (via controlled phase jumps on the Raman lasers) and fitted interference fringes by packets of 20 points to extract the $1\sigma$ statistical uncertainties from a data set of 100 packets.
The fitted visibility data pinpoints the mirror's angle with an uncertainty of 4.0 (2.3) steps in the $z$ ($y$) direction, corresponding to an angular uncertainty of 1.5 (0.9)~$\mu$rad (assuming an average conversion factor for both mirror directions of $0.38~\mu$rad/step).

The data reported in Fig.~\ref{fig:contrast_ZY} may also be interpreted in position space: the difference in direction between the two effective Raman wave-vectors translates into a spatial separation between the two wave packets at the last pulse  of the interferometer, given by $\delta\vec{r}=2\frac{\hbar}{M}(\vec{k}\stxt{B}-\vec{k}\stxt{T}) = 4v_{R}T\delta\theta \etheta$, where $v_{R}\simeq 3.5$~mm.s$^{-1}$ is the one-photon recoil velocity of the Cesium atom. We  expect a  loss of interferometer contrast as this separation becomes comparable to the coherence length  of the atomic wavepacket, $L_{\rm{coh}}$. Fitting the curves in Fig.~\ref{fig:contrast_ZY} with a gaussian model, we extract a standard deviation  in the horizontal direction of $\sigma_{\theta}\simeq 14.1(0.9) \ \mu$rad, which determines (as a result of the convolution between two gaussian wavepackets) the value of the coherence length  $L_{\rm{coh}}=2\sqrt{2}v_{R}T\sigma_{\theta}\simeq 56(4)$~nm (the values are similar in the vertical direction).
The comparison between the two interpretations (velocity and position)  shows that the wavepacket does not saturate the Heisenberg uncertainty relation, as $L_{\rm{coh}} \times m\sigma_v \simeq 2.6\times \frac{\hbar}{2}$ (relative uncertainty below $10\%$).

\subsection{Determination of the optimal operational velocity}
\begin{figure*}
	\centering
	\includegraphics[width=\linewidth]{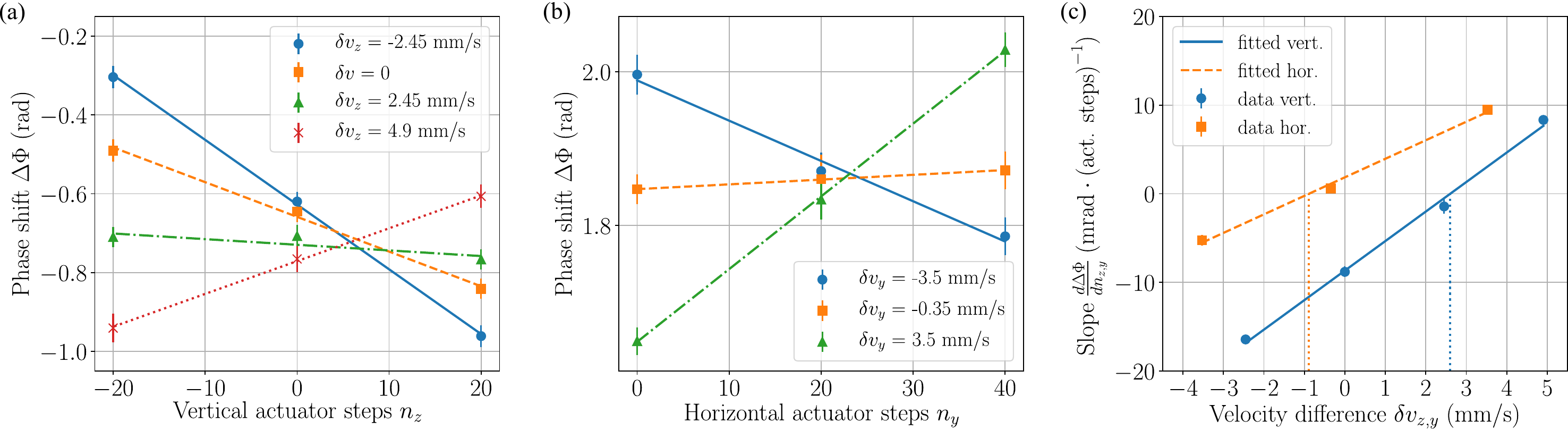}
	\caption{Induced phase shift as a function of mirror angular position (in actuator steps) in (a) vertical and (b) horizontal directions, for different values of velocity offsets ($\delta v_{z,y}$). Blue dots (solid blue line), orange squares (dashed orange line), green triangles (dashed-dotted green line) and red crosses (dotted red line) data (linear fit) in (a) are obtained with $\delta t = 0.25, 0, -0.25, -0.5$~ms, respectively; blue dots  (solid blue line), orange squares (dashed orange line) and green triangles (dashed-dotted green line) data (linear fit) in (b) are obtained with $\delta\beta_{y} = -0.7, -0.07, 0.7$~mrad, respectively (see text). The error bars are standard Allan deviation of phase in a half-an-hour-long acquisition. (c) Slopes (phase shift per actuator step) extracted from the fits of the data shown in panels (a) - (blue dots) and (b) - orange squares. Solid blue and dashed orange lines are the linear fits to corresponding data. The vertical dotted lines indicate the values of optimal velocities that minimize the bias of the atom interferometer.}
	\label{fig:ZYdirection_and_v-slope}
\end{figure*}

From Eqs.~\eqref{eq:probability_to_phase_link} and \eqref{eq:velocity_theta_deltaPhi}, the dependence of the transition probability can be explicitly written as
\begin{equation}
P(\vec{v}, \delta\theta\etheta) = \frac{1}{2} + C\cos\big(2T \keff (\delta\theta_z \delta v_z + \delta\theta_y \delta v_y) +\phi \big),
\label{eq:probability_to_phase_link2}
\end{equation}
with $\delta v_z = v_z +gT$ and $\delta v_y=v_y$ respectively being the vertical and  horizontal components of the offset of the mean velocity vector from the optimal one.
To find the optimal velocity, we measure the phase shift of the interferometer when varying the velocity of the atoms and the mirror alignment.
We  first introduce a controlled velocity shift along the vertical ($z$) direction by shifting in time the four pulses by an equal amount $\delta t$ with respect to their initial values, which is equivalent to varying the initial mean vertical velocity of the atoms by $\delta v_{z}= -g\delta t$.
Fig.~\ref{fig:ZYdirection_and_v-slope} (a) shows the phase shift measured for three mirror positions and four different values of induced $\delta v_{z}$. The optimal position of the mirror $n_{z0}$ corresponds to the crossing point of the lines, which we extract from simultaneous fitting of four data sets ($i=1..4$) with linear functions having common offset parameters: $a_{i}(n_{z}-n_{z0}) + b$. We find the value of $n_{z0}=6.1(6)$, slightly different from the value obtained from optimization on the contrast signal. 
This method improves the determination of the optimum mirror position with respect to the contrast measurements, as it yields an uncertainty of 0.6 steps, i.e. about $0.2 \ \mu$rad.

To induce a controlled variation of the horizontal launch velocity in the reference frame of the sensor, we tilt the whole experiment in the $(yz)$-plane by an angle $\delta\beta_{y}$, thus changing the projection of the total acceleration $\vec{g}$ on the y-axis~\cite{NoteTilt}. The resulting deviation of the mean horizontal velocity becomes $\delta v_{y} = g\delta\beta_{y}(t_{1}+T)$.
In Figure~\ref{fig:ZYdirection_and_v-slope} (b) we show the recorded phase shift at different values of mirror tilt along y-axis, for three different values of $\delta v_{y}$ achieved with corresponding sensor tilts $\delta\beta_{y}$. With the same simultaneous fitting routine as used for z-direction, we extract the value of $n_{y0}=23.1(5)$ that reveals a small difference of ($\approx1.3~\mu\textrm{rad}$) as compared to optimization with the contrast curve (Fig.~\ref{fig:contrast_ZY}(b)).

We now combine the results for  both directions and plot in Figure~\ref{fig:ZYdirection_and_v-slope}(c) the corresponding phase shifts per actuator step as a function of velocity variation. The expected dependence is a linear function intercepting the coordinates' origin. We fit the data with $\frac{d\Delta\Phi}{d n_{z}} (\textrm{mrad/act. steps})= -8.7(5)+3.33(16)\delta v_{z}\textrm{(mm/s)}$ and $\frac{d\Delta\Phi}{d n_{y}} (\textrm{mrad/act. steps}) = 1.9(4)+2.09(13)\delta v_{y}\textrm{(mm/s)}$ for z- and y-directions, respectively.
This representation of the data reveals the value of velocities to be used in order to minimize the phase shift with respect to mirror misalignment:  2.60(19)~mm/s for the z-direction and -0.89(18)~mm/s for the y-direction.
With this method, we are therefore able to minimize the interferometer bias by finding the optimal velocity with an accuracy of about 0.2~mm.s$^{-1}$ in both relevant directions. Together with the determination of the optimum mirror position with an accuracy of about $0.2~\mu$rad, we can thus constrain the phase bias of the interferometer to 0.5~mrad, which corresponds to a rotation rate bias of the gyroscope of $1\times 10^{-10}$~rad.s$^{-1}$.

\subsection{Comparison with the theoretical phase shift}

Extracting the slopes of the data of Fig.~\ref{fig:ZYdirection_and_v-slope}(c) provides a calibration of the step-to-angle conversion factor of the piezo-motorized  mirror mount in both directions, by matching the measurements with the expected slope of $2T\keff=11.80~\textrm{mrad}\cdot\mu\textrm{rad}^{-1}\cdot\textrm{(mm/s)}^{-1}$ given by Eq.~\eqref{eq:general_deltaPhi}. From these data we obtain conversion factors of $0.282(14) \ \mu\textrm{rad}/ \textrm{step}$ and $0.177(11) \ \mu\textrm{rad}/ \textrm{step}$ for the vertical and horizontal directions, respectively. Both numbers significantly deviate from the expectation of $0.385~\mu\textrm{rad}/ \textrm{step}$.
Despite the fact that exact values of the conversion factor are not crucial to tune the interferometer at the optimal operating point, we explain here the  reasons for the discrepancy (and give additional details in appendix \ref{sec:appendix_model_slope}). In the following, we include corrections arising from the finite size of the atomic cloud after expansion which becomes comparable to the size of  the detection region, as well as the size of the interrogating Raman  beam at the last light pulse. 

For the typical temperature of our atomic cloud of $T_{at}=1.8(3)~\mu\textrm{K}$, the  thermal rms velocity of $3.0(2)~v_R$ drives an isotropic expansion resulting in a gaussian width of the cloud of  $\sigma_{x,y,z}\simeq 10$~mm after 984~ms of time of flight (time of detection). We take into account two contributions.
\textit{(i)} First, we consider the detection-related correction: the detection region for each of the lightsheets  resembles a square in horizontal plane with a 30~mm side. The atoms falling outside this square in $x$ and $y$-directions are not detected, which corresponds to excluding about $17\%$ of the atoms in the wings of Gaussian distribution along each horizontal axis, assuming the cloud centered in the detection region in the horizontal plane. The $z$-axis is not affected in this scenario as all the atoms pass sequentially in the detection region. 
\textit{(ii)} Second, for both axes ($y$ and $z$), we account for the exact shape of the intensity profile of the bottom Raman beam that performs the last $\pi/2$ pulse of the interferometer. The associated intensity variation gives rise to a spatially inhomogeneous beamsplitter efficiency, which can be parametrized by a weighting function $w(z)=2\sqrt{P(y,z)(1-P(y,z))}$, where $P(y,z)$ is the local probability of the Raman transition. As a result, the contribution of the atoms to the interference signal depends  on their position inside the cloud.

As the velocity distribution is mapped on a spatial distribution after ballistic expansion, the cut by the detection and the inhomogeneous beamsplitter efficiency modifies the contributions of atoms of different velocity groups. This  can be  interpreted as a shift of the initially induced mean velocity towards an effective mean velocity. For example, in the $y$-direction, the atomic cloud becomes shifted at the detection region proportionally to the tilt-induced acceleration by $\delta y = g\delta\beta_{y}(t_{1}+t_{\textrm{det}})^2 /2$, which amounts to 3.5~mm for $\delta\beta_{y}=0.7$~mrad. The faster side of the cloud becomes subsequently more cut than the slower side, which results in a slower induced effective mean velocity.

\begin{figure}[t]
	\centering
	\includegraphics[width=\linewidth]{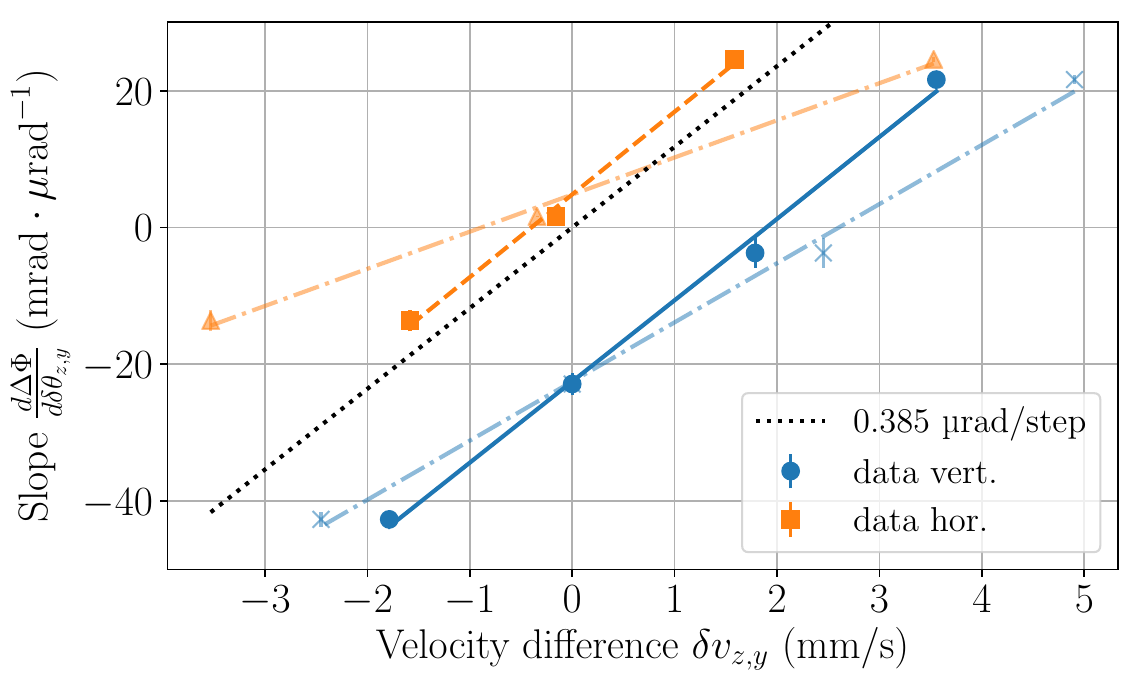}
	\caption{Phase shift per $\mu$rad of angular mirrors' mismatch. Blue dots (orange squares) data points correspond to the data sets for vertical (horizontal) directions in Figure~\ref{fig:ZYdirection_and_v-slope}(c), with the corrected effective velocities extracted from the model presented in the main text. Solid blue (dashed orange) lines are linear fits to the data of vertical (horizontal) directions. Semi-transparent blue crosses (orange triangles) show the same data of vertical (horizontal) direction, but with original velocity values, with the dashed-dotted lines being corresponding linear fits (to guide the eye). The dotted black line is the theoretical expectation for $0.385~\mu$rad/step scaling factor.}
	\label{fig:v-slope}
\end{figure}

We apply this model to obtain the corrected, effective, mean velocities for both $z$- and $y$-direction. In Figure~\ref{fig:v-slope}, we compare the original (Fig.~\ref{fig:ZYdirection_and_v-slope}) and corrected datasets. The effective initial velocities in the $y$-direction are almost a factor of 2 smaller as compared to the tilt-induced ones, which explains the previously observed strong discrepancy. The resulting fitted slope changes from the original value of $5.43(33)~\textrm{mrad}\cdot\mu\textrm{rad}^{-1}\cdot\textrm{(mm/s)}^{-1}$ to the value of $12.10(72)~\textrm{mrad}\cdot\mu\textrm{rad}^{-1}\cdot\textrm{(mm/s)}^{-1}$. 
In the $z$-direction, we find smaller velocity corrections that shift the slope from $8.65(41)~\textrm{mrad}\cdot\mu\textrm{rad}^{-1}\cdot\textrm{(mm/s)}^{-1}$ to  $11.90(59)~\textrm{mrad}\cdot\mu\textrm{rad}^{-1}\cdot\textrm{(mm/s)}^{-1}$. The corrected slopes in both directions match well the expectation. 
We note, however, that this correction is  sensitive to the exact knowledge on several parameters including cloud temperature, size of detection region, Gaussian waist and full size of the interrogating beam. For example, a $10\%$ relative error in the cloud temperature  generates  $5\%$ ($8\%$) variation of the slope value in the $z$ ($y$) direction. Additional experiments supporting our   model are presented in appendix \ref{sec:appendix_model_slope}.

\section{Point Source Analysis}
\label{sec:point_source}

In case of a relatively large mirror misalignments, the  contrast of the interferometer becomes significantly reduced as a result of destructive averaging of the velocity-dependent phase shifts  (see Fig.~\ref{fig:contrast_ZY} for $\sim 80$ actuator steps). The total phase shift can  thus be hardly extracted from the integrated atomic signal. However, as the  velocity distribution of the atoms closely resembles  that of a point source after our  long expansion time, the initial velocity of each atom is directly mapped onto its position in the detection region. This opens the possibility to resolve  velocity-dependent phase shifts, as it has been done in previous studies by means of imaging techniques with a camera~\cite{Dickerson2013,Hoth2016}. Here we apply this technique, known as point-source interferometry (PSI),  in the case of fluorescence detection.
\begin{figure}[t]
	\centering
	\includegraphics[width=\linewidth]{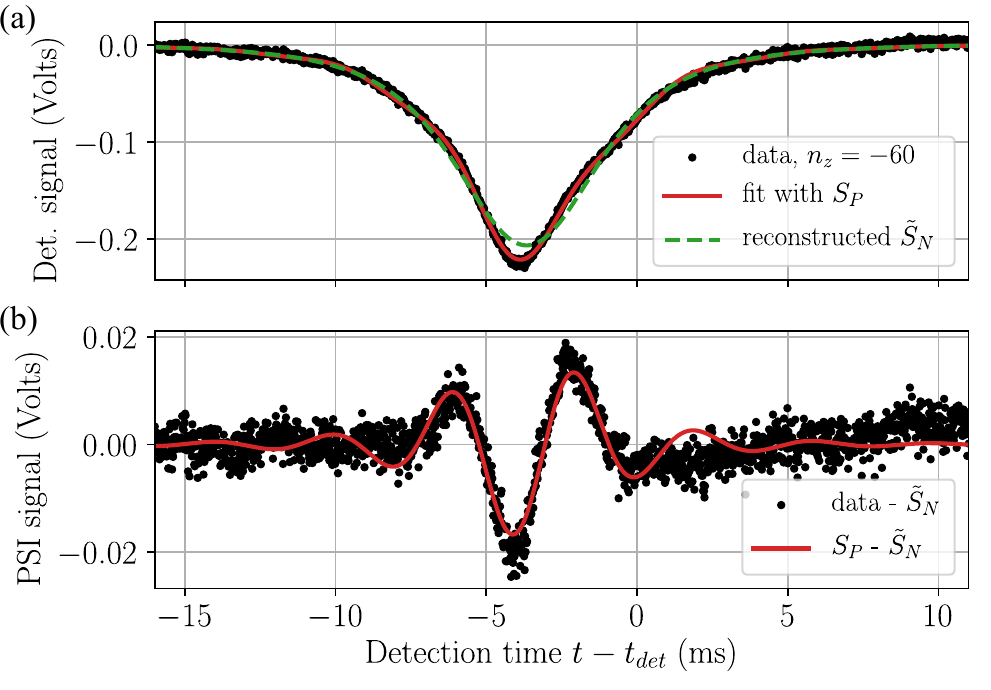}
	\caption{(a) Time-of-flight signal in the top light sheet fitted with the function $S_{P}(t)$ (solid red line), and corresponding reconstructed total atom number profile $\tilde{S}_{N}(t)$ (dashed green line). (b) Difference between the data and the reconstructed total atom number profile and extracted PSI waveform $S_{P}(t)-\tilde{S}_{N}(t)$ (solid red line).}
	\label{fig:probability_fitted_and_dTheta_vs_steps}
\end{figure}

As explained in section \ref{subsec:detection_system}, our detection scheme does not discriminate the position of the atom within the light sheets such that all velocity classes along $x$- and $y$-axes contribute to the signal. We thus model the detected signals with time-of-flight profiles for probability ($S_{P}(t)$) and total atom number ($S_{N}(t)$) as
\begin{align}
\label{eq:model_detection}
\begin{split}
\displaystyle S_{P}(t)=&\int\limits_{-\infty}^{\infty}\int\limits_{-\infty}^{\infty}dv_{x}dv_{y}\int\limits_{\bar{v}_{z}(t)- d/2t_{det}}^{\bar{v}_{z}(t)+d/2t_{det}}dv_{z} f(\vec{v})P(\vec{v}, \delta\theta\etheta),
\\
S_{N}(t)=&\int\limits_{-\infty}^{\infty}\int\limits_{-\infty}^{\infty}dv_{x}dv_{y}\int\limits_{\bar{v}_{z}(t)-d/2t_{det}}^{\bar{v}_{z}(t)+d/2t_{det}}dv_{z} f(\vec{v}),
\end{split}
\end{align}
with
\begin{equation}
\bar{v}_{z}(t) = v_{z0} + \left(g-\frac{v_{z0}}{t_{det}}\right)(t-t_{det}),
\label{eq:time-to-velocity_mapping}
\end{equation}
where $\bar{v}_{z}(t)$ is the initial vertical velocity of an atom that arrives at the center of the detection region at time $t$, and $v_{z0}$ is the center of the launched velocity distribution~\cite{footnoteModel}. 
 The integration goes over the initial velocities of the atoms governed by the    distribution function $f(\vec{v})=f(v_x,v_y,v_z)$. The details of the exact form of distribution chosen to model the data are given in appendix \ref{sec:appendix_vel_distrib}.
In the limit of $\delta\theta_{y,z}\rightarrow 0$ the phase shift in Eq.~\eqref{eq:velocity_theta_deltaPhi} becomes velocity-independent, and we recover the simple proportionality relations for the integrated signals considered so far:
\begin{equation}
\int S_{P}(t)dt\propto N_{F=4}  \ \ \ \text{and} \ \ \ \int S(t)dt\propto N.
\end{equation}

In  Figure \ref{fig:probability_fitted_and_dTheta_vs_steps}(a) we show a typical time-of-flight profile for a mirror relative misalignment  $\delta\theta_{z}\approx 25~\mu$rad, as recorded  from the upper detection light sheet (see Fig.~\ref{fig:Figure_detection_scheme}). We fit this data with the function $S_{P}(t)$ defined in Eq.~\eqref{eq:model_detection} and parameterized as follows: $A\times S_{P}(t; v_{z0}, \bar{v}_{c}, B, C, \delta\theta_{z}, \phi) + D$ (solid red line). The free parameters are the amplitude, $A$, the offset $D$, the fringe contrast $C$, the mean phase $\phi$, the mirror misalignment $\delta\theta_z$.  The free parameters $\{\bar{v}_{c}, B, v_{z0}\}$  characterize the Lorentz-B velocity distribution defined by:
\begin{equation}
f(v_z)=\frac{f(0)}{\left(1+\left(\frac{v_{z}-v_{z0}}{\bar{v}_{c}}\right)^2\right)^B}
\label{eq:lorBdistrib}
\end{equation}
(see Appendix~\ref{sec:appendix_vel_distrib} for more details).
Importantly, in this data we minimize the angular misalignment along the $y$-direction, such that the phase shift (Eq.~\eqref{eq:probability_to_phase_link2}) becomes $v_{y}$-independent. Using the fitted parameters of the  velocity distribution, we reconstruct the average time-of-flight profile $\tilde{S}_{N}$ (dashed green line). The deviation of the data with respect to  $\tilde{S}_{N}$ represents thus a bare contribution due to the interference, as illustrated in  Figure \ref{fig:probability_fitted_and_dTheta_vs_steps}(b).

\begin{figure}[t]
	\centering
	\includegraphics[width=\linewidth]{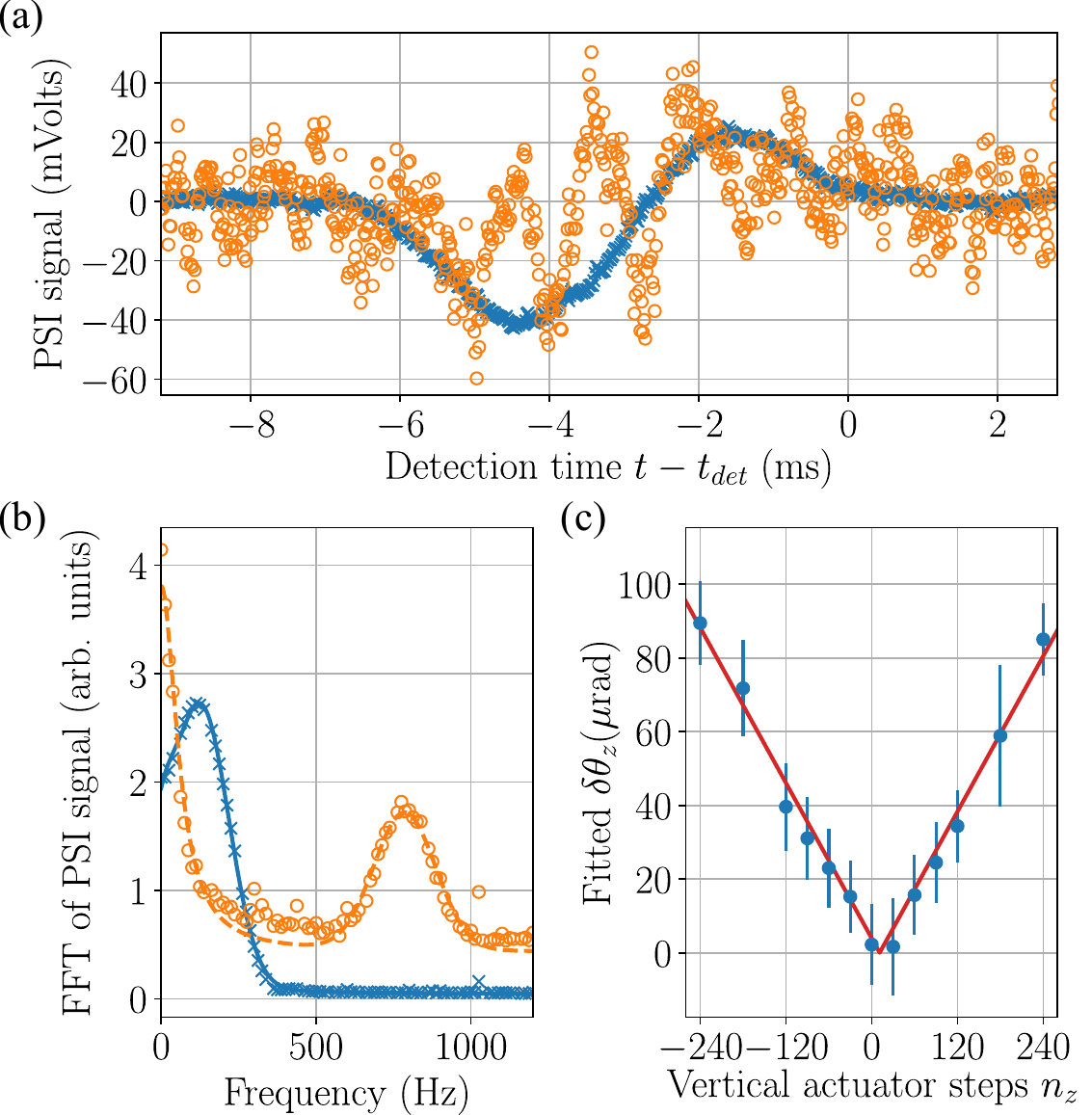}
	\caption{(a) Typical PSI signals for the two values of actuator steps $n_z = 30$ (blue crosses) and $n_z = 240$ (orange circles). (b) Fourier transform of the data in panel (a): $n_z = 30$ (blue crosses) and $n_z = 240$ (orange circles). Solid blue and dashed orange lines are the fits of corresponding data. In (a) and (b), the data for  $n_z = 240$ are enhanced $\times10$ for clarity. (c) Vertical angular misalignment for a range of actuator step values, extracted from the fits as shown in panel (b). Plotted values (error bars) correspond to the fitted center position (1-$\sigma$ width). Red line is the fit to the resulting data array.}
	\label{fig:PSI_and_dTheta_vs_steps}
\end{figure}

Applying this method to a broad range of mirrors' relative misalignment $\delta\theta_z$ encounters, however, a number of limitations. The robustness of the fitting routine is highly dependent on the exact model of the velocity distribution of the atoms, since an associated error translates into a spatial modulation that can mimic the fringes. Additionally, while using this method in the case of small angular misalignment ($n_{z}\lesssim 60$), we observe strong couplings between different fit parameters. Here, the fringes' period becomes so large that it exceeds the size of the cloud, making the variation in $\delta\theta_{z}$ coupled with variations of fringe contrast $C$ and Lorentz-B parameters. Lastly, the technical imperfections as shot-to-shot fluctuations in atom number and small defects of the time-of-flight profiles render fits unstable as the PSI signal of interest is much smaller than the bulk envelope.

In order to robustly capture the mirrors' relative misalignment $\delta\theta_z$ in a broad range, we thus adopt a different method based on Fourier analysis of the PSI signal. We start by recording 96 identical probability traces (as in Figure~\ref{fig:probability_fitted_and_dTheta_vs_steps}(a)) for the actuator step values in the range of $[-240, 240]$. For a given value of $n_z$, we then take the mean of 96 traces, which serves a reference atomic profile as the PSI phase is sufficiently randomized (see Eq.~\ref{eq:probability_to_phase_link}). We subtract this reference profile from each probability trace in order to obtain a bare PSI signal. The corresponding profiles for the values of $n_{z}=-30$ and $n_{z}=-240$ are shown in Figure~\ref{fig:PSI_and_dTheta_vs_steps}(a). Here, we can clearly observe the PSI waveform extending over multiple periods for the large values of $n_z$, which highlights the potential for tracking large $\delta\theta_z$ angles with PSI analysis as compared to integrating the entire fluorescence trace. To retrieve the value of $\delta\theta_z$, we perform a fast Fourier transform (FFT) for each of the 96 PSI profiles and take the mean of the obtained FFT traces for better signal-to-noise ratio. The resulting mean FFT profiles of $n_{z}=-30$ and $n_{z}=-240$ data are shown in Figure~\ref{fig:PSI_and_dTheta_vs_steps}(b). We fit these profiles with phenomenological function which is a sum of Lorentzian profile centered at 0~Hz (accounting for background) and Gaussian profile (signal of interest). Based on Eqs.~\ref{eq:probability_to_phase_link2} and \ref{eq:time-to-velocity_mapping}, from the fitted center of the Gaussian profile we extract the value of $\delta\theta_z$. As uncertainty in $\delta\theta_z$, we take  the fitted width of the Gaussian peak, as one may expect from the finite frequency resolution linked with the total length of the cloud envelope.

Figure~\ref{fig:PSI_and_dTheta_vs_steps}(c) shows the results for the fitted angular mirror misalignment, revealing the expected linear behavior. The extracted values for $\delta\theta_{z}$ are in good agreement with the expectation. Fitting the full data with an absolute linear function gives $\delta\theta_{z}(\mu\rm{rad}) = 0.352(2)~\rm{steps}/\mu\rm{rad}\times |n_{z}-11(9)|$. The conversion factor between the actuator steps and the angular pitch in the vertical direction is quite close to the value obtained in section \ref{sec:results}, as for the proof-of-concept demonstration shown here. A full quantitative analysis would require a more elaborate fitting of the time-of-flight profiles, which goes beyond this study.

\section{Conclusion}
\label{sec:conclusion}

To summarize, we presented a method to minimize the systematic effect associated with the misalignment between the retro-reflecting Raman mirrors coupled to the atom trajectory  in a 4 light pulse atom interferometer operated in fountain configuration.  We showed that the relative misalignment between the mirrors can be zeroed  within an accuracy of $0.2 \ \mu\text{rad}$ and that the optimal velocity can be set within an accuracy of  $0.2$~mm.s$^{-1}$. The resulting bias represents a phase shift of 0.5~mrad, which amounts to a rotation rate bias of $1\times 10^{-10}$~rad.s$^{-1}$.
With this level of adjustment, achieving a rotation rate stability of $1\times 10^{-11}$~rad.s$^{-1}$ translates in ensuring a stability of the mean velocity (in $y$ and $z$ directions) at the level of $20\ \mu$m.s$^{-1}$ and a stability of the mirror relative misalignment at the level of $20$~nrad. While the former is already reached in our experiment, stabilizing the angular misalignments will require  dedicated mechanical engineering and/or active temperature stabilization to reach the desired level.

Using the point source interferometry technique with fluorescence detection, we also showed how large ($\sim 90 \ \mu$rad) mirror relative misalignments in the vertical direction can be extracted even when the contrast from the integrated transition probability signal vanishes. At larger misalignments, the contrast of the PSI traces diminishes due to the finite velocity resolution of our detection light sheets ($10$~mm.s$^{-1}$). Using an imaging system  with a camera would allow to extend the method to the 2 relevant directions. Therefore, the PSI method appears as a good starting point for a rough alignment of the mirrors, before using the higher-precision step associated with the introduction of controlled velocity shifts.

The method which we presented here is not specific to a  gyroscope architecture but will serve other atom interferometric sensors based on spatially separated interrogation laser beams. In the case of large momentum transfer (LMT) atom optics, the effect will scale with the diffraction order.
As an example, proposals of ground-based gravitational wave detectors in the 0.1 to 10~Hz frequency band shall employ a similar interferometer configuration as that presented here, but with atom interferometric sensors spatially distributed along a common laser baseline in a gradiometer setup \cite{Canuel2018,Canuel2019elgar}. In that case, differential phase noise between distant sensors will occur from the uncorrelated atomic velocities from one sensor to the other. With LMT orders of 1000 (e.g. 500 times $2\hbar k$ momentum transfers), and assuming that the mirror relative alignment can be zeroed with an accuracy of 10~nrad, reaching phase noise levels of the order of 1~$\mu$rad.Hz$^{-1/2}$ in the desired frequency band (to reach strain sensitivities $\sim 10^{-22}$~Hz$^{-1/2}$ \cite{Canuel2019elgar}) will require a control of the atomic velocities at the level of $10$~nm.s$^{-1}$.Hz$^{-1/2}$.

\section*{Acknowledgments}

We acknowledge the financial support from Ville de Paris (project HSENS-MWGRAV), FIRST-TF (ANR-10-LABX-48-01),   Centre National d'Etudes Saptiales (CNES), Sorbonne Universit\'es (project SU-16-R-EMR-30, LORINVACC) and Agence Nationale pour la Recherche (project PIMAI, ANR-18-CE47-0002-01). L.A.S. was funded by Conseil Scientifique de l'Observatoire de Paris, D.S. by Direction G\'en\'erale de l'Armement, and M.A. and R. Gautier by the EDPIF doctoral school. We thank Enrique Morell for his work on a part of the optical system used to complete this study.

\appendix

\section{Derivation of the systematic effect}
\label{sec:AppendixA}
\renewcommand{\thefigure}{A\arabic{figure}}
\setcounter{figure}{0}

We give here the details of the derivation of the systematic shift Eq.~\eqref{eq:velocity_theta_deltaPhi} resulting from the angular misalignment between the two Raman beams coupled to the trajectory of the atoms.
\begin{figure}[b]
	\centering
	\includegraphics[width=.8\linewidth]{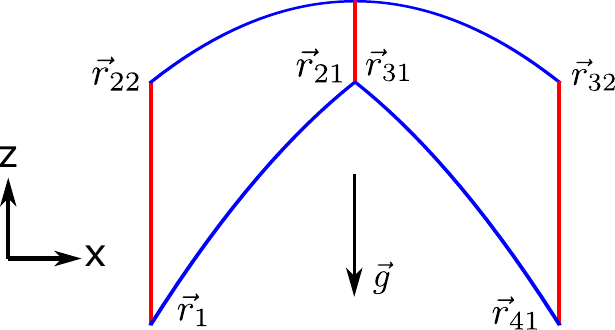}
	\caption{Schematic of the wavepacket propagation (classical trajectories) in the $(xz)$ plane (not to scale) with the notations for the positions used in the calculation.}
	\label{fig:4Pspace_rij}
\end{figure}
We first determine the classical position of the atom at the Raman pulses. We write as $\vec{v}_0$ ($\vec{r}_1$) the velocity (position) of the atom wave packet at the first pulse, and $\vec{r}_{ij}$ the position of the atom at pulse $i$ in the arm $j$ of the interferometer, with $j=1$ being the path corresponding to a diffraction at the first pulse (see Fig.~\ref{fig:4Pspace_rij}). For simplicity, we will place the coordinate's origin at $\vec{r}_1$:
\begin{align}
\vec{r}_{1}  & = 0 \\ \nonumber
\vec{r}_{21} & = \frac{T}{2}\vec{v}_0 + \frac{T^2}{8}\vec{g} + \frac{1}{2}\poskeff\vec{e}_r \\ \nonumber
\vec{r}_{22} & =\frac{T}{2}\vec{v}_0 + \frac{T^2}{8}\vec{g} \\ \nonumber
\vec{r}_{31} & = \frac{3T}{2}\vec{v}_0 + \frac{9T^2}{8}\vec{g} + \poskeff\left[\left(\frac{3}{2}-\cos\delta\theta\right)\vec{e}_r - \sin\delta\theta\vec{e}_\theta\right] \\ \nonumber
\vec{r}_{32} & = \frac{3T}{2}\vec{v}_0 + \frac{9T^2}{8}\vec{g} + \poskeff(\cos\delta\theta\vec{e}_r + \sin\delta\theta\vec{e}_\theta) \\ \nonumber
\vec{r}_{41} & = 2T\vec{v}_0 + 2T^2\vec{g} + \poskeff\left[(2-\cos\delta\theta)\vec{e}_r - \sin\delta\theta\vec{e}_\theta\right]
\end{align}

The phase of the interferometer reads:
\begin{align}
\Delta\Phi & = (\vec{k}\stxt{B}\vec{r}_1 - \vec{k}\stxt{T}\vec{r}_{21} + \vec{k}\stxt{T}\vec{r}_{31} - \vec{k}\stxt{B}\vec{r}_{41}) - (\vec{k}\stxt{T}\vec{r}_{22} - \vec{k}\stxt{T}\vec{r}_{32})\\ \nonumber
& = 2T\keff(\vec{v}_0 + T\vec{g})[(\cos\delta\theta -1)\vec{e}_r + \sin\delta\theta\vec{e}_\theta] \\ \nonumber
& + 2\phikeff(\cos\delta\theta -1)
\end{align}
To the first order in $\delta\theta$, we obtain:
\begin{equation} \label{eqn:phaseDesalignV0}
\Delta\Phi = 2T\keff\delta\theta\vec{e}_\theta.(\vec{v}_0 + T\vec{g}).
\end{equation}

\section{Additional support  for the comparison with the theoretical phase shift}
\label{sec:appendix_model_slope}
\renewcommand{\thefigure}{B\arabic{figure}}
\setcounter{figure}{0}

In order to further support our velocity-selectivity-based correction model, we acquire an additional data for the vertical direction, similar to the dataset in Figure~\ref{fig:ZYdirection_and_v-slope}(a) but with a  different experimental arrangement. First, we replace the bottom Gaussian-beam collimator with a collimator having a rectangular (top-hat) intensity profile with full width of 30~mm and same total optical power \cite{Mielec2018}. Second, we record now the full detection trace that allows us to select different parts of the cloud in the $z$-direction in post-processing and to determine the corresponding  phase shifts. 
\begin{figure}[t]
	\centering
	\includegraphics[width=\linewidth]{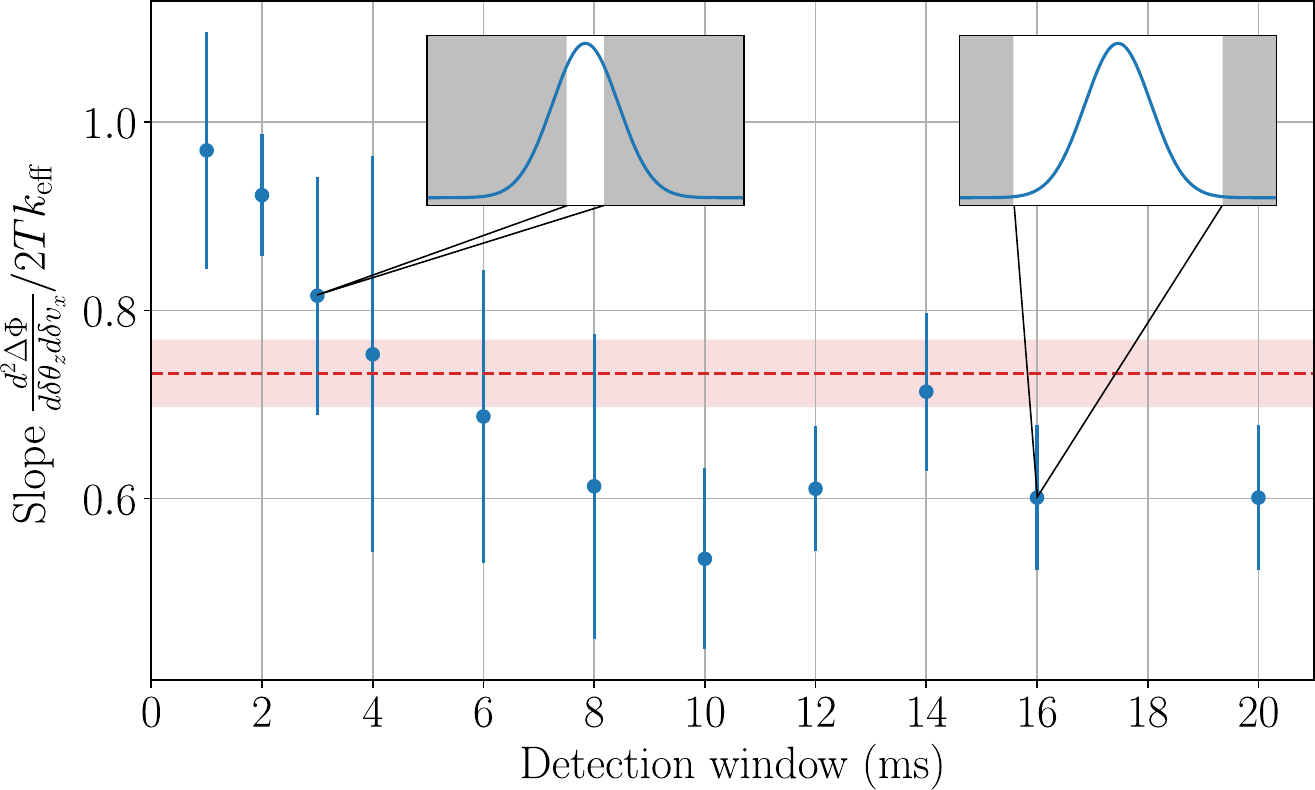}
	\caption{Normalized vertical slope as a function of detection window size. Red dashed horizontal line (pink-shaded region) corresponds to the fit (fit error bar) to the data of Fig.\ref{fig:ZYdirection_and_v-slope}(c) (i.e. the limit of using the full detection window). The non-shaded regions in the insets represent the parts of the detection signal used for data processing.}
	\label{fig:scaling_det_window}
\end{figure}

In Figure~\ref{fig:scaling_det_window} we show the dependence of the slope $\frac{d^2 \Delta\Phi}{d\delta\theta_{z}d\delta v_{z}}$ normalized to the expected slope of $2T\keff$ on the size of the detection window centered around the maximum of the signal. As the detection window size approaches zero, the value of the normalized slope tends to unity, thus well recovering the expected scaling. At large values of detection window (more than half of integrated signal inside), we find a slope slightly below the previously measured value (corresponding  to an actuator conversion factor of $0.282 \ \mu\textrm{rad}/ \textrm{step}$). This shift is most likely attributed to small quantitative difference between the Gaussian and the top-hat collimators. A more advanced modeling of this experimental result goes beyond this work.

\section{velocity distribution}
\label{sec:appendix_vel_distrib}
\renewcommand{\thefigure}{C\arabic{figure}}
\setcounter{figure}{0}
\begin{figure}[b]
	\centering
	\includegraphics[width=\linewidth]{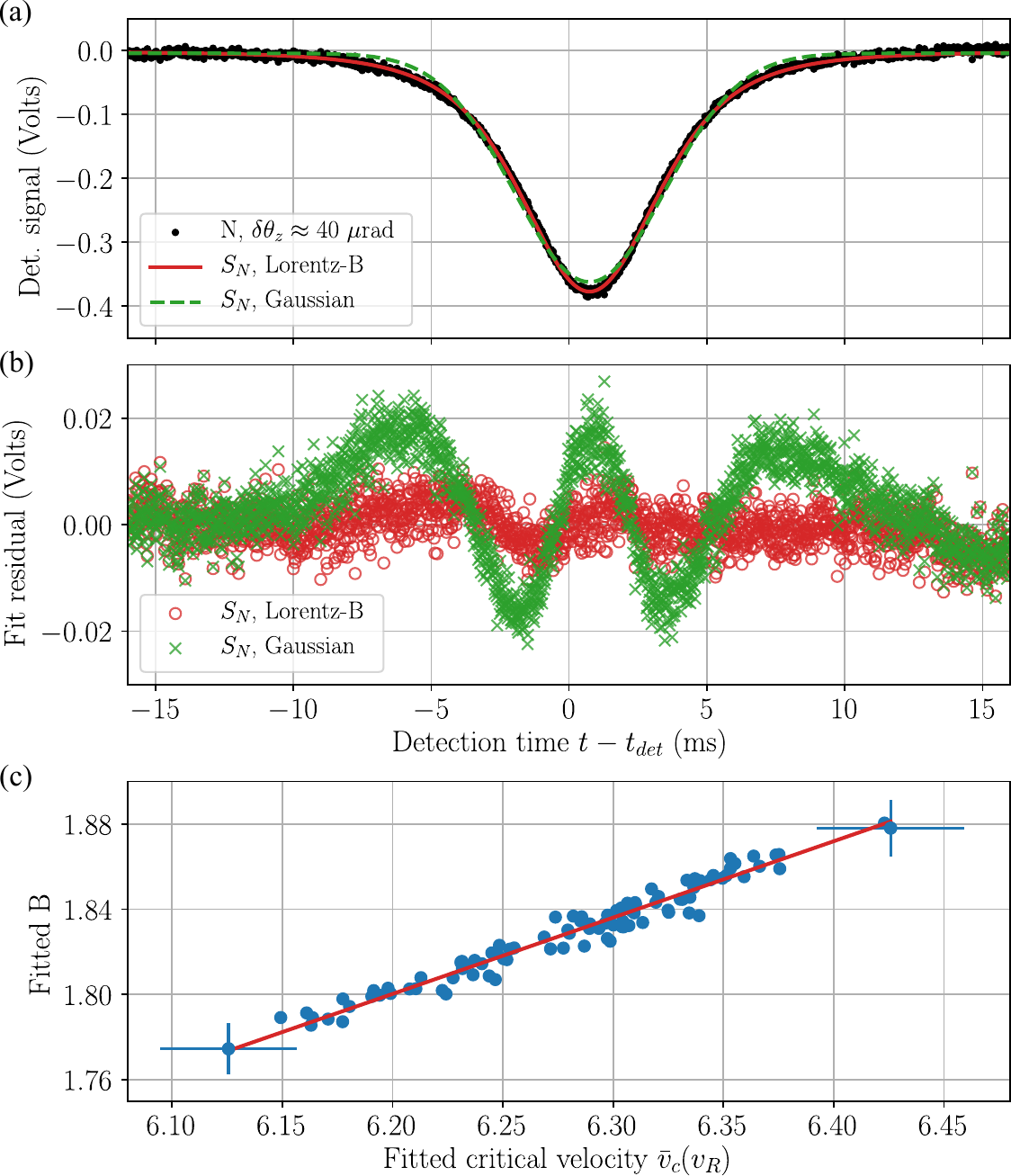}
	\caption{(a) Typical atom number signal (black dots) fitted with $S_{N}$ function using Lorentz-B (solid red line) and Gaussian (dashed green line) velocity profiles for detection region $d=10$~mm, (b) residuals of the Lorentz-B (empty red circles) and Gaussian (green crosses) fits from panel a, (c) parameters of the Lorentz-B profile extracted from 100 identical experimental shots, fitted with linear dependence using orthogonal distance regression routine: $B=0.358(8) \bar{v}_{c}/v_R - 0.42(4)$. For clarity, typical error bars of the single points are shown only at extremities of the data array.}
	\label{fig:velocity_distribution}
\end{figure}
Understanding the velocity distribution in optical molasses is itself a challenging task that was a subject of intense research~\cite{Hodapp1995tds} and resulted in the  development of a 1D-model based on solution of the Fokker-Planck equation for simultaneous diffusion in momentum and configuration space. Following this work, we expect the atomic velocity distribution for our experimental parameters to have a quasi-thermal-equilibrium profile given by so-called Lorentz-B function~\cite{Hodapp1995tds}:
\begin{equation}
f(v)=\frac{f(0)}{(1+(v/\bar{v}_{c})^2)^B},
\label{eq:9}
\end{equation}
where parameter $B$ is linked to the critical velocity parameter $\bar{v}_c$ with $B=\frac{1}{3\sqrt{3}} \frac{\bar{v}_c}{v_R}$, for the case of a model atom considered in Ref.~\cite{Hodapp1995tds}. 
The analytical derivation of the  velocity distribution in case of 3D optical molasses as well as the generalization of the 1D model to a 3D case is more complex (see, e.g. Ref~\cite{SortaisPhDthesis} page 172). We thus aim here at an empirical characterization of our velocity distribution in the vertical direction only, by analyzing time-of-flight profiles. 

In Figure~\ref{fig:velocity_distribution}, we show our results of fitting the fluorescence trace for the total atomic signal  (as the one shown in Fig.~\ref{fig:Figure_detection_scheme}(b), bottom). The integrated (along $x$- and $y$-directions) profiles are fitted with two $S_{N}$ functions as previously introduced in Eq.~\ref{eq:model_detection}, with $f(v)$ distributions corresponding to Lorentz-B and Gaussian one-dimensional profiles (Fig.~\ref{fig:velocity_distribution}(a). The strong fit residuals (Fig.~\ref{fig:velocity_distribution}(a) illustrate significant deviation of the actual velocity distribution from the thermal  Gaussian one. The Lorentz-B profile fits the data much closer. Therefore, when we refer to the temperature of the cloud we rather mean the effective temperature - the value extracted from the best-fitted Gaussian profile, typically about $1.8~\mu\textrm{K}$.

We further analyze a single series of 100 identical atom number pictures and extract the corresponding array of $B, \bar{v}_{c}$ pairs (Fig.~\ref{fig:velocity_distribution}(c)). The statistical shot-to-shot fluctuation of fitted parameters suggest about $20\%$ peak-to-peak variation of the cloud temperature, and a clear linear scaling between $B$ and $\bar{v}_{c}$. We use three similar datasets of 100 points each, taken with time difference of about 30 minutes, in order to have an estimate of parameters' variation over longer time. We obtain the relation: $B=0.367(11) \bar{v}_{c}/v_R - 0.45(7)$, for typical 1-hour timescale. 
As compared to the 1D model, we therefore find a factor of about 2 larger slope and a significant offset value.

\bibliography{trajectory_paper}

\begin{thebibliography}{23}
\bibitem [{foo({\natexlab{a}})}]{footnoteMirror}%
  \BibitemOpen
  \href@noop {} {} ({\natexlab{a}}),\ \bibinfo {note} {the step size of the
  linear actuator (N-470 by PI) depends on its load, on the direction of the
  translation and on the total number of steps performed since the start of
  operation. Based on the data sheet, we initially estimated a translation of
  20(1)~nm per motor step, with an error bar due to hysteresis. Given the
  60.5~mm baseline of our mirror mount, this translated in angular variation of
  0.33(2)~$\mu$rad/step, while our direct angular calibration yielded
  0.39(2)~$\mu$rad/step and 0.38(2)~$\mu$rad/step for vertical and horizontal
  direction, respectively.}\BibitemShut {Stop}%
\bibitem [{Not()}]{NoteTilt}%
  \BibitemOpen
  \href@noop {} {}\bibinfo {note} {The tilt is stabilized on the signal from
  auxiliary tilt-meter, with two actuators independently controlling tilt
  angles $\delta\beta_{x,y}$ in two orthogonal vertical planes XZ and
  YZ.}\BibitemShut {Stop}%
\end{thebibliography}
\label{biblio}

\end{document}